\documentclass[12pt]{article}

\usepackage{gensymb}
\usepackage{url}
\usepackage{amssymb,amsmath, amsfonts}
\usepackage{cases}
\usepackage{latexsym}
\usepackage{graphicx}
\usepackage{relsize}
\usepackage{natbib}
\usepackage{rotating}
\usepackage[includeheadfoot,left=1.in,right=1.in,top=1.in,bottom=1.in,bindingoffset=0.in,nohead]{geometry}
\usepackage[onehalfspacing]{setspace}
\usepackage{booktabs}
\usepackage{appendix}   
\usepackage[pdfborder={0 0 0}]{hyperref}
\usepackage{float}
\usepackage{xcolor}
\usepackage{caption}
\usepackage{subcaption}
\usepackage{minibox}
\usepackage{floatpag}
\usepackage[most]{tcolorbox}

\begin{document}
	
	\title{A Benchmark   Model\\for  Fixed-Target Arctic  Sea Ice Forecasting}

	\author{Francis X. Diebold\\University of Pennsylvania \and Maximilian G\"obel\\University of Lisbon \\$~$}
	
	\maketitle

\bigskip
\bigskip
%
%
%
%

\bigskip

\bigskip
	
	\begin{spacing}{1}
		
		\noindent \textbf{Abstract}: 
			We propose a  reduced-form benchmark predictive model (BPM) for fixed-target forecasting of Arctic sea ice extent, and we provide a case study of its real-time performance for target date September 2020.   We visually detail the evolution of the statistically-optimal point, interval, and density forecasts as time passes, new information arrives, and the end of September approaches. Comparison to the BPM  may prove useful for evaluating and selecting among various more sophisticated  dynamical sea ice models, which are widely used to quantify the likely future evolution of Arctic conditions and their two-way interaction with economic activity.
				
		\thispagestyle{empty}
				
		\bigskip
		\bigskip
		\bigskip
		\bigskip
		\bigskip	
		\bigskip
		\bigskip
				\bigskip
				\bigskip
				\bigskip

				\noindent	\textbf{Replication files} (data, R code, matlab code, etc.): Available in the ancillary materials repository at   \url{https://arxiv.org/abs/2101.10359}. 
				
				\bigskip

		\noindent {\bf Acknowledgments}:  For helpful communications we thank, without implicating, Uma Bhatt, Philippe Goulet Coulombe, Walt Meier, Glenn Rudebusch, and Boyuan Zhang. 
		
		\bigskip
		
		
		\bigskip
		
		\bigskip
		
		\noindent {\bf Contact}: fdiebold@sas.upenn.edu

	\end{spacing}
	
%
	
	\clearpage
	
	\setcounter{page}{1}
	\thispagestyle{empty}

\section{Introduction}

The Arctic is warming much faster than the rest of the planet, and it has emerged as a crucial focal point of climate change study.  The path and pattern of Arctic sea ice diminution is of particular interest, and sea ice forecasting has received significant attention.  From a real-time online perspective, there are two key forecast types: fixed-horizon (e.g., each month we might forecast one month ahead, month after month, ongoing) and fixed-target (e.g., each month we forecast a fixed future target month, month after month, ending when we arrive at the target month).  In this paper we consider the fixed-target scenario, which has generated  substantial interest in highlighting Arctic sea ice diminution both within years (as September 30 is approached, say) and across years (comparing the sequence of Septembers, say).\footnote{A third forecast type arises from an offline perspective -- the so-called extrapolation forecast, with  a fixed origin and an expanding range of horizons, as with a forecast for every month from now until the end of the century.}  

For example,  each summer since 2008 the Sea Ice Prediction Network (SIPN)  has sponsored the Sea Ice Outlook  (SIO) competition for fixed-target prediction of September average daily Arctic sea ice extent.\footnote{See \url{https://www.arcus.org/sipn} for SIPN, and see \url{https://www.arcus.org/sipn/sea-ice-outlook} for SIO.} September extent forecasts are produced by many research groups mid-month in June, July, and August, and evaluated once September ends and the outcome is known.  Insightful post-season SIO assessments have been   produced annually (the most recent is \cite{SIOinterim2020}), and similarly-insightful multi-year retrospective SIO assessments have been  produced occasionally \citep{StroeveEtAl2014,HamiltonStroeve2016,Hamilton2020}.  Those assessments focus primarily on the forecasting skill of the SIO point-forecast ensembles.  

In this paper we take an approach different from the SIO analyses, drilling very far down, focusing not on a point-forecast ensemble but rather on the  point, interval, and density forecast paths for a single and very simple model (which we call the Benchmark Predictive Model, or BPM) in a single season (2020). The broad insights gained --  associated in particular  with the evolution of forecast uncertainty from a simple yet sophisticated reduced-form sea ice forecasting model as time progresses  and the target date is approached -- are of  wide use.  Indeed the BPM approach and results feature prominently in the  ``glide chart" climate model evaluation and comparison framework developed  in \cite{DGGC2022}, in which the BPM is used as the ``naive" reference model in climate model skill scores. 

We proceed as follows.  In section \ref{two}  we introduce the target-date  forecasting framework and the BPM.  In section \ref{five} we provide the  2020 case study.  We consider forecasts made on the SIO dates, as well as a generalized set of forecasts made daily from June through September, and we pay particular attention to forecast uncertainty as the target date is approached.  We conclude in section \ref{six}.

\section{A Benchmark Predictive Model for Arctic Sea Ice Extent} \label{two}

We  consider target-date forecasting for  September average daily sea ice extent, $SIE_{9}$, conditioning on the expanding historical sample as we move from June through the end of September.  We forecast using a simple reduced-form model, which we call the ``benchmark predictive  model" (BPM), regressing  September extent on four covariates:
\begin{equation}
 \label{full}
 SIE_{9} ~{\rightarrow} ~c, ~ Time, ~SIE_{LastMonth}, ~SIE_{ThisMonthSoFar}, ~ SIE_{Today},
  \end{equation}
where $SIE_p$ denotes average daily extent during period $p$ (hence, for example, $SIE_9$ denotes September extent), ``$\rightarrow$" denotes ``is regressed on", and the rest of the notation is obvious.  

Approximately following SIO, we make $SIE_{9}$ forecasts on four days: 6/10, 7/10, 8/10 and 9/10.\footnote{We include a 9/10 forecast even though the 2020 SIO did not.  The 9/10 forecast is of interest because September average extent is not known with certainty until the \textit{last} day of September, well after 9/10.  Indeed subsequent  installments of the SIO will solicit 9/10 forecasts.}  Immediately,  the 6/10 regression used to produce the June forecast is
 $$
 SIE_{9} ~{\rightarrow}~ c, ~ Time, ~ SIE_{5}, ~ SIE_{6/1\_ 6/10} ,~ SIE_{6/10},
 $$
the 7/10 regression used to produce the July forecast is
$$
SIE_{9} ~{\rightarrow}~ c, ~ Time, ~ SIE_{6}, ~ SIE_{7/1\_ 7/10} ,~ SIE_{7/10},
$$
 the 8/10 regression  used to produce the August forecast is
$$
SIE_{9} ~{\rightarrow}~ c, ~ Time, ~ SIE_{7}, ~ SIE_{8/1\_ 8/10} ,~ SIE_{8/10},
$$
and the 9/10 regression  used to produce the September forecast is
$$
SIE_{9} ~{\rightarrow}~ c, ~ Time, ~ SIE_{8}, ~ SIE_{9/1\_ 9/10} ,~ SIE_{9/10}.
$$

 Perhaps surprisingly given their simplicity, the  BPM  forecasts  are quite sophisticated in certain respects of relevance for forecasting Arctic sea ice:
 
\begin{enumerate}
	
	\item They capture low-frequency trend dynamics, via conditioning on $Time$.
	
	\item   They capture medium-frequency inertial (autoregressive) dynamics around  trend, via conditioning on  $SIE_{LastMonth}$.

\item They capture high-frequency dynamics by augmenting the conditioning on historical monthly information (via $SIE_{LastMonth}$) with potentially-invaluable recent daily  information, via $SIE_{ThisMonthSoFar, t}$ and $SIE_{Today, t}$.  

\item  They readily enable probabilistic quantification of forecast uncertainty, which  lets us move easily from point forecasts to interval and density forecasts.

\item They are based on  a BPM estimated using direct rather than iterated projections.\footnote{One makes a multi-period ``iterated" forecast with a one-period-ahead estimated model, iterated forward for the desired number of periods.  In contrast,  one makes a  multi-period ``direct" forecast with a horizon-specific multi-period-ahead estimated model.} Direct projections are theoretically superior under model misspecification (which is always the relevant case), because they directly minimize the relevant multi-step predictive loss.\footnote{See \cite{Ing2003}, Theorem 4 and Corollary 3.}

 \item  They are easily made day-by-day, using model parameter estimates optimized day-by-day to the remaining predictive horizon, thanks to the BPM's simplicity. We exploit this fact below  to make and examine not only the  monthly SIO forecasts, but also 120 daily forecasts from June through September.

\end{enumerate}

 \noindent The BPM combination of trivial simplicity and subtle sophistication makes it an appropriate benchmark for skill score comparisons, as in \cite{DGGC2022}. On the one hand, one would hope that a best-practice scientific model (e.g., a sophisticated structural climate model) should outperform the simple BPM, but on the other hand, it may not be easy!


\section{Forecasting 2020 September Arctic Sea Ice Extent} \label{five}

\subsection{Estimation }

\begin{table}[tb] 
	\begin{center}
		\caption{September 2020 Arctic Sea Ice Extent:  Regression Results and Forecasts }
		\label{forecs1}
		\begin{tabular}{l cccccc}
			\toprule 
			&  & June 10 &  July 10 & Aug 10  & Sept 10 \\
			\midrule 
			\addlinespace[10pt]
			$c$          							&   & -2.75  & -2.87 & -1.83    & -0.77  \\
			$Time$       							&  & -0.04  & -0.02 & -0.003 &  0.003  \\
			$SIE_{LastMonth}$   			&  & -0.13    & 0.18   &  0.25   &   0.39 \\
			$SIE_{ThisMonthSoFar}$ 	& {} & -1.94 &  -0.61 &   -0.45 &  -0.28  \\
			$SIE_{Today}$ 					& {} & 2.93   & 1.38 &  1.26     &  0.97 \\  
			\midrule
				$ \hat{\sigma}$           &    &  0.462   & 0.403 & 0.267 &  0.100    \\
		 $\bar{R}^2$ &      &   0.83     &  0.87 & 0.94 &   0.99     \\
			\midrule 
			$\hat{\mu}$	(Sept. point forecast)                &     &  4.32    & 3.84   &  4.34   &   {3.93}     \\
			$\hat{\mu} \pm 2  \hat{\sigma}$  	(Sept. interval forecast)  &  & [3.40,5.25]  & [3.03,4.65]  & [3.80,4.87] &  [{3.73} ,{4.13}]  \\
						\midrule \addlinespace[10pt]
			Sept realization: {3.92}  &  &   &  & & \\
			\bottomrule 
		\end{tabular}
	\end{center}
	\begin{singlespace} \footnotesize
				Notes: The left-hand-side variable in each of the four regression models is September extent (monthly average of daily values).  The estimation samples have 41 annual observations, 1979-2019.  Our daily extent measure is the National Snow and Ice Data Center (NSIDC) Sea Ice Index, Version 3 (\url{https://doi.org/10.7265/N5K072F8}).  Until August 1986, data are reported only every other day, and we fill missing days observations with the average of the two adjacent days.  Forecasts are made on the 10$^{th}$ of each month, on June 10 using the estimated June model,  on July 10 using the estimated July model, and so on through September 10 using the estimated September model. The point forecast is $\hat{\mu}$, and the interval forecast is 	$\hat{\mu} \pm 2  \hat{\sigma}$.  See text for details.  
		\end{singlespace}
\end{table}

 The left-hand-side variable of the BPM is September extent.  September 2020 extent data were obviously unavailable on June 10, July 10, August 10, or September 10.  Hence all estimation samples are  1979-2019, for a total of 41 annual observations.\footnote{Our daily extent measure is the National Snow and Ice Data Center (NSIDC) Sea Ice Index, Version 3 (\url{https://doi.org/10.7265/N5K072F8}),  which  uses the NASA team algorithm to convert microwave brightness readings into ice coverage \citep{Fettereretal2017}.  Until August 1986, data are reported only every other day, and we fill missing days with the average of the two adjacent days.} 
 
 Estimation results appear in the top and middle panels of Table \ref{forecs1}.  Several points are worth noting.  First, 
the negative linear trend becomes progressively less important as September approaches, whereas the positive autoregressive effect $SIE_{LastMonth}$ becomes  progressively \textit{more} important as September approaches.  This is completely natural. The conditioning on May extent in the June 10 forecast, for example, is of little value for forecasting September extent, so the trend plays an important role.  In contrast, moving to the end of the summer, the conditioning on August extent in the September 10 forecast is of great value for forecasting September extent, so the trend plays almost no role.

Second,  $SIE_{ThisMonthSoFar}$ has a negative effect and $SIE_{Today}$ has a positive effect.  Hence the estimates, and the forecasts that we construct from them, are influenced not  just by $SIE_{Today}$, but also  by  $SIE_{Today}$ 	\textit{relative} to  $SIE_{ThisMonthSoFar}$.

Finally, adjusted R-squared ($\bar{R}^2$) naturally increases toward 1.0 as September approaches, because the value of the conditioning information ($SIE_{LastMonth}$, $SIE_{ThisMonthSoFar}$, $SIE_{Today}$) increases as September approaches. In parallel, the standard error of the regression ($\hat{\sigma}$) naturally decreases toward 0 as September approaches, again because the value of the conditioning information increases as September approaches.

\subsection{Forecasting }

To use an estimated forecasting model to make a point forecast, we simply insert  the relevant right-hand-side variables, all of which are known at the time the forecast is made.  For example, to form the July 10 forecast we evaluate the fitted July model at $Time{=}42$, $SIE_{LastMonth}{=}SIE_{6/2020}$,   $SIE_{ThisMonthSoFar}{=} SIE_{7/1/2020\_ 7/10/2020}$,  $SIE_{Today}{=} SIE_{7/10/2020}$.\footnote{There is typically a 1-day data availability lag, so we would actually insert  $SIE_{LastMonth}{=}$ $SIE_{6/2020}$,  $SIE_{ThisMonthSoFar}{=}$ $SIE_{7/1/2020\_ 7/9/2020}$,  $SIE_{Today}{=}$ $SIE_{7/9/2020}$.} This point forecast is an estimate of the mean of $SIE_{7/2020}$ conditional on $Time{=}42$, \\ $SIE_{LastMonth}{=} SIE_{6/2020}$,   $SIE_{ThisMonthSoFar}{=} SIE_{7/1/2020\_ 7/10/2020}$,  and $SIE_{Today}{=} SIE_{7/10/2020}$. Hence we denote the point forecast by $\hat{\mu}$ in Table \ref{forecs1}.  

Now consider interval forecasts (predictive intervals). Let us stay with the same July example.
 To make an interval forecast we need an estimate of the standard deviation of $SIE_{7/2020}$ conditional on the same covariates:  $Time{=}42$, $SIE_{LastMonth}{=}SIE_{6/2020}$,   \\ $SIE_{ThisMonthSoFar}{=}$  $SIE_{7/1/2020\_ 7/10/2020}$, and $SIE_{Today}{=}SIE_{7/10/2020}$.  The standard error of the regression, denoted $\hat{\sigma}$ in Table  \ref{forecs1}, is precisely such an estimate.\footnote{Note that $\hat{\sigma}$ measures true forecast uncertainty, which is a very different concept from the cross-section dispersion in the ensemble of forecasts, $\hat{d}$. We want $\hat{\sigma}$, and in general $\hat{\sigma} {\ne} \hat{d}$.}  An interval forecast (ignoring parameter estimation uncertainty) is then $\hat{\mu} {\pm} 2 \hat{\sigma}$.  If the regression disturbances are approximately Gaussian, then the $\hat{\mu} {\pm} 2 \hat{\sigma}$ interval is an approximate 95\% predictive interval.\footnote{One could use simulation-based bootstrap procedures to accommodate parameter estimation uncertainty and/or non-Gaussian disturbances in forming interval forecasts, but we do not pursue that here.}

Finally, again ignoring parameter estimation uncertainty, consider density forecasts (predictive densities). If the regression disturbances are approximately Gaussian, then the full predictive density is approximately $N(\hat{\mu}, \hat{\sigma}^2)$.\footnote{As with the interval forecast case, bootstrap procedures could be used to accommodate parameter estimation uncertainty and/or non-Gaussian disturbances.}

\subsubsection{Four Month-by-Month Predictive Densities} \label{4forecs}

\begin{figure}[t]
	\begin{center}
		\caption{Arctic Sea Ice Extent: Four Predictive Densities for September 2020} \label{GlidingDens_SEP_T1}
		\includegraphics[trim={0mm 0mm 0mm 50mm},clip,scale=0.2]{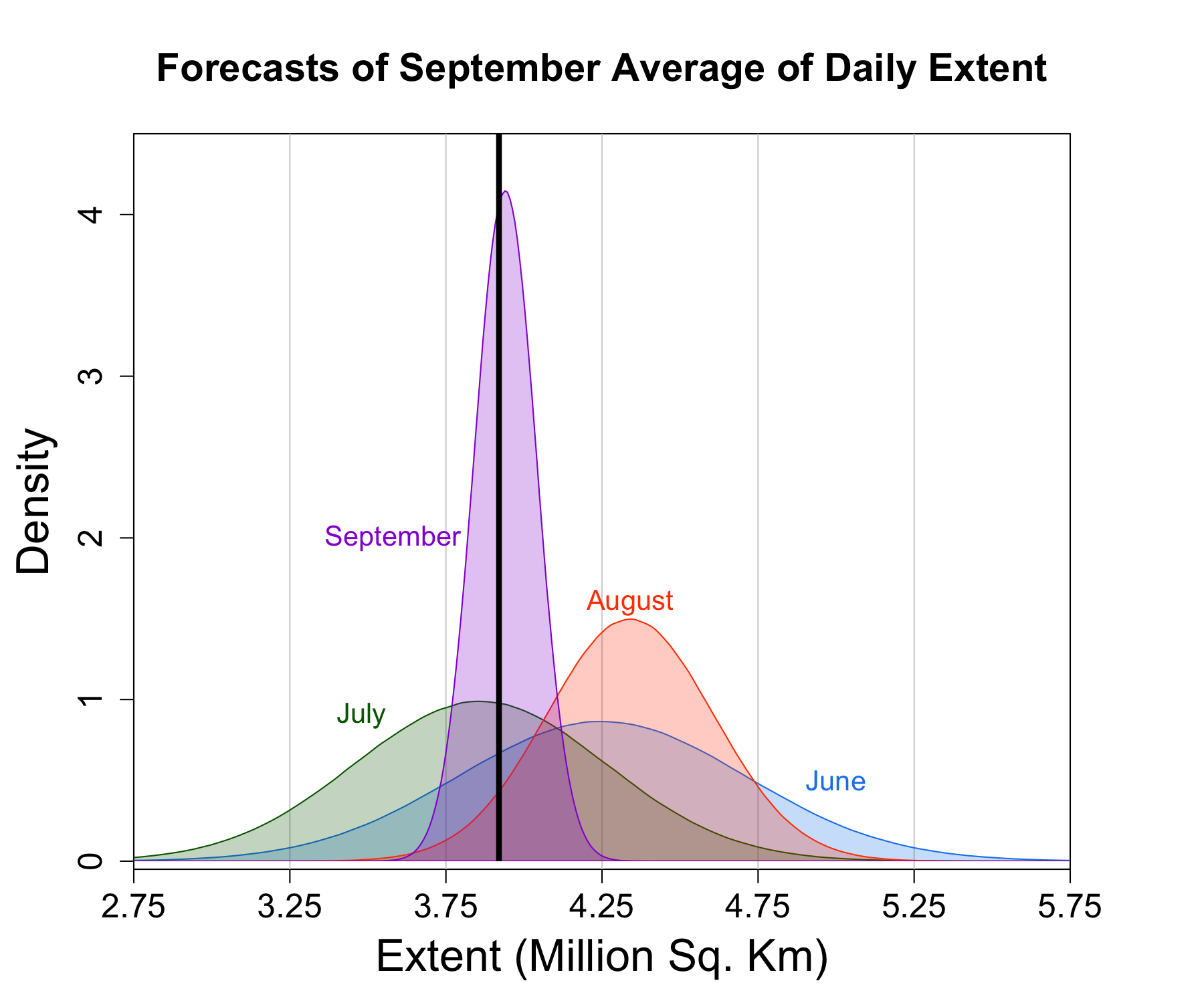}\\
	\end{center}
	\begin{spacing}{1.0}  \footnotesize \noindent Notes: We show four predictive densities for September 2020 Arctic sea ice extent (monthly average of daily values). Forecasts are made on the 10th of the  month.  The vertical black line is the  realized September value. 	See text for details.
	\end{spacing}
\end{figure}

In Figure \ref{GlidingDens_SEP_T1} we show the four monthly predictive densities (June, July, August, and September) corresponding to our generalized SIO exercise that includes a September 10 forecast.  The density locations (their means, the $\hat{\mu}$'s	in Table \ref{forecs1})  naturally evolve throughout the summer as the conditioning information evolves, but they eventually get closer to the end-of-September value.  The density mean is above the realization in June, below in July, above again in August, and then  almost spot-on in  September. 

  Not unrelated, and importantly, the forecast uncertainty as captured by the predictive density dispersion ($\hat{\sigma}$ in Table \ref{forecs1}) decreases monotonically moving through the summer: from Table \ref{forecs1} it is $0.46$, $0.40$, $0.27$, $0.10$ for June, July, August, and September, respectively.

\subsubsection{120 Day-by-Day Predictive Densities} \label{120}

\begin{figure}[p]  
	\begin{center}
		\caption{Arctic Sea Ice Extent: Day-by-Day Predictive Densities for September 2020} \label{3D_Plot_SEP_V4}
		\includegraphics[trim={0mm 0mm 0mm 5mm},clip,scale=0.75]{{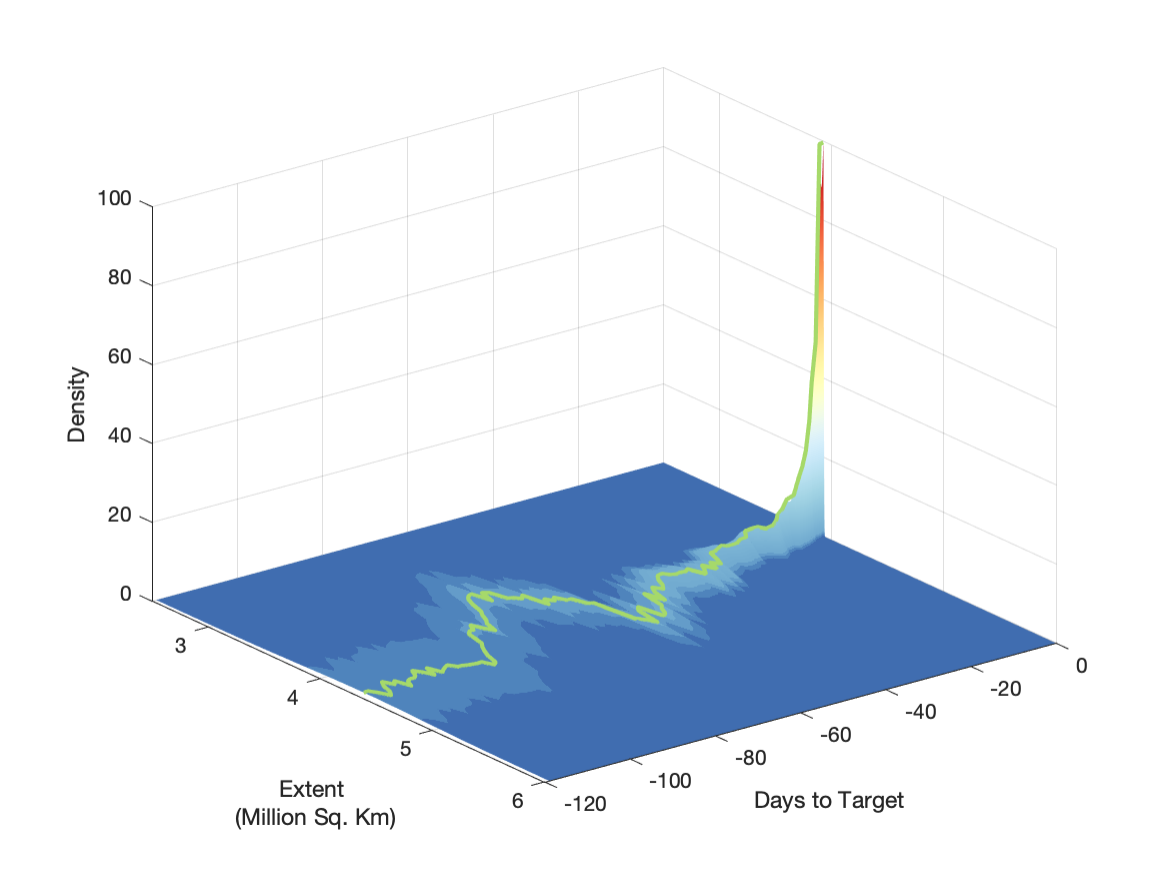}}\\
		\includegraphics[trim={10mm 0mm 10mm 10mm},clip,scale=0.75]{{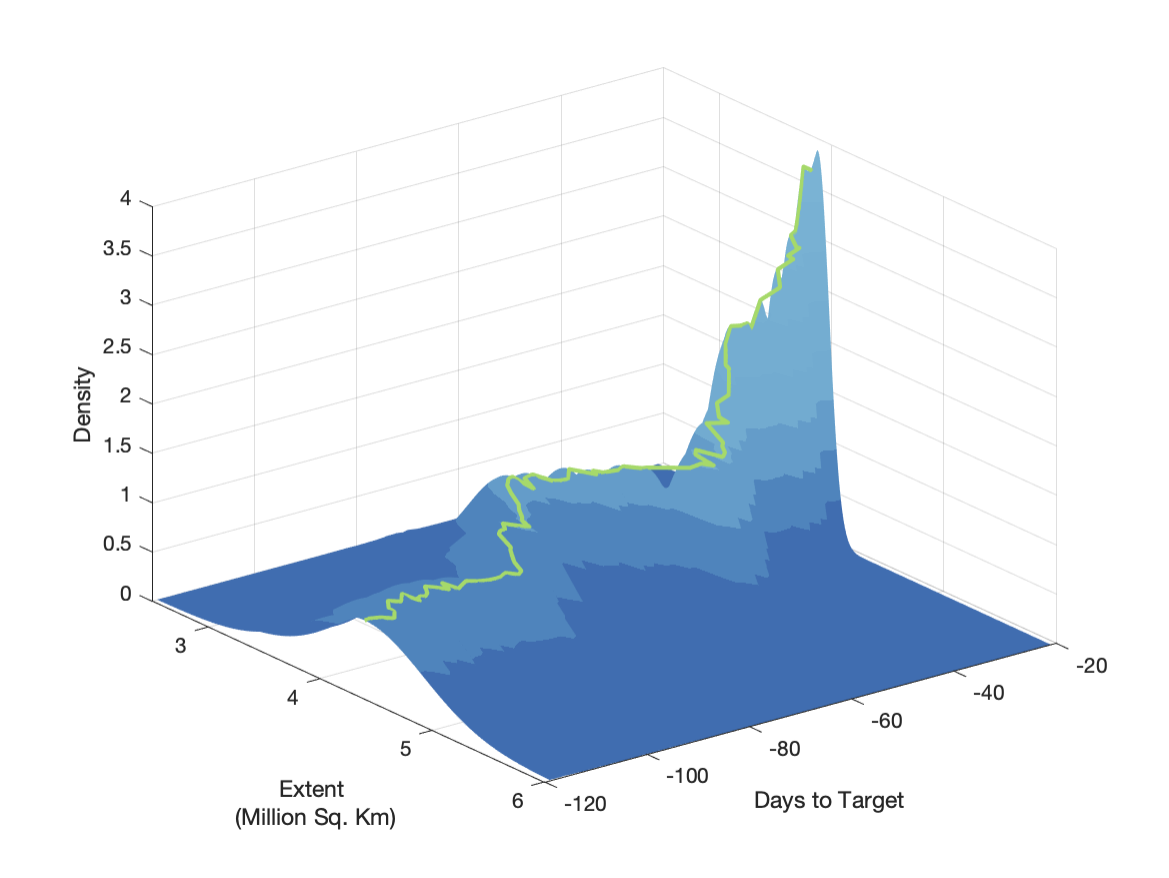}}
	\end{center}
	\begin{spacing}{1.0}  \footnotesize \noindent Notes: We show the sequence of 120 day-by-day predictive densities for September 2020 average daily Arctic sea ice extent as September 30 is approached. The horizontal axis represents the number of days until the end of September, and the green line is  the point forecast (the mean of the predictive density). Top: we show the entire [-120, 0] sequence.  Bottom: we zoom in on [-120, -20], to enhance visualization detail.  
	\end{spacing}
\end{figure}

There is nothing sacrosanct about the set of once-per-month SIO forecast dates examined thus far.  Given the simplicity of our forecasting model and its estimation, we can examine many other dates. We simply generalize the BPM from 
\begin{equation}
	SIE_{9} ~{\rightarrow} ~c, ~ Time, ~SIE_{LastMonth}, ~SIE_{ThisMonthSoFar}, ~ SIE_{Today}
\end{equation}
to
\begin{equation}
	\label{full2}
	SIE_{9} ~{\rightarrow} ~c, ~ Time, ~SIE_{LastMonth}, ~SIE_{Last30Days}, ~ SIE_{Today},
\end{equation}
and the framework is otherwise unchanged.

In Figure \ref{3D_Plot_SEP_V4}, we show predictive densities for the 120 days leading  to the end of September, produced using 120 different estimated models. In the top panel we plot the entire sequence [-120, 0], and in the bottom panel we plot only [-120, -20] to enhance visualization detail.  Throughout, the horizontal axis represents the number of days until the end of September, and the green line is the evolving point forecast (the mean of the predictive density).  One can readily see the densities wandering left and right as new information arrives, but nevertheless eventually rising sharply and clustering tightly around the realized value as the end of September nears.

\begin{figure}[t]  
	\begin{center}
		\caption{Arctic Sea Ice Extent: Day-by-Day Predictive Intervals for September 2020} \label{GlidePoint_SEP_2020}
		\includegraphics[trim={0mm 30mm 0mm 50mm},clip,scale=0.25]{{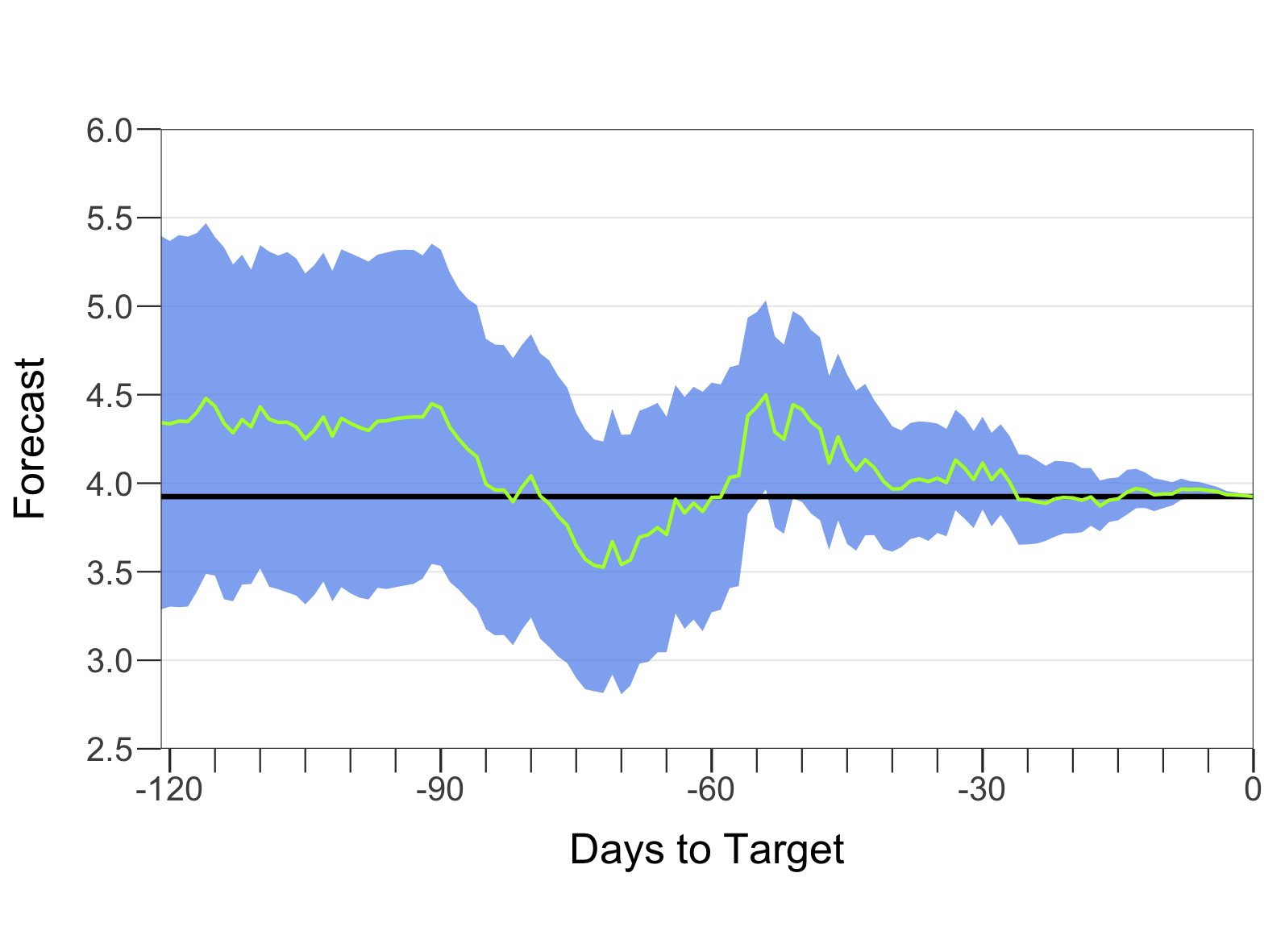}}\\
	\end{center}
	\begin{spacing}{1.0}  \footnotesize \noindent Notes: We show the sequence of day-by-day prediction intervals for September 2020 average daily Arctic sea ice extent as September 30 is approached. The horizontal axis represents the number of days until the end of September, the green line is  the point forecast (the midpoint of the prediction interval), and the shaded area is the  $\pm 2$ standard error band. The horizontal black line is the  realized September value.
	\end{spacing}
\end{figure}

In Figure \ref{GlidePoint_SEP_2020} we reduce the predictive densities  to predictive intervals. As the target date approaches, the interval forecast midpoint (the point forecast, $\hat{\mu}$) evolves as the conditioning information evolves, converging to the eventually-realized September value.  Simultaneously the interval forecast width ($4 \hat{\sigma}$) also evolves as information accumulates, converging  to zero by the target date.\footnote{Of course the densities of Figure \ref{3D_Plot_SEP_V4} and the intervals of Figure \ref{GlidePoint_SEP_2020} are isomorphic in a Gaussian environment -- if one knows the density, then one knows the interval, and conversely, so that nothing new is  learned by reduction of densities to  intervals.  Nevertheless the sequence of  intervals may be visually revealing in certain ways that the sequence of  densities is not, more clearly emphasizing both the point forecast trajectory and its associated uncertainty, and hence serving as a complement rather than a substitute for the sequence of  densities.  Moreover, and importantly, the environment may \textit{not} be Gaussian, in which case the $\pm 2 \hat{\sigma}$ intervals are still a useful and transparent  quantification of forecast uncertainty, even if they lose their interpretation at 95\% confidence intervals.}

%

\section{Concluding Remarks}  \label{six}

We earlier asserted that our benchmark predictive model (BPM) is quite sophisticated in certain respects.  As it turned out, its performance in the 2020 Sea Ice Outlook competition was in the middle of the pack, a thoroughly respectable performance for a simple BPM.  And the point, of course, is not that the BPM should dominate its competitors, but rather that it should serve as a simple benchmark against which allegedly  more sophisticated competitors can be compared.

Following that path, one may  use the BPM as the reference model in  ``skill score glide charts" for climate model evaluation and comparison, tracking relative forecasting performance of competitors vs BPM as time evolves and the target date is approached.  Such competitor vs BPM skill score glide charts are proposed and explored in work in progress \citep{DGGC2022}.

Skill score competitors may include more sophisticated reduced-form models, including, for example, models that:

\begin{enumerate}

\item incorporate nonlinearity, whether parametrically (e.g.,  \cite{DRice}), or nonparametrically as in a variety of statistical machine learning methods (e.g., \cite{hastie2009});

\item  incorporate  and forecast  the  entire daily sea ice extent history (note that we do not model the entire daily history -- we model the monthly history augmented  with certain aspects of the very recent daily history);

\item drop the normality assumption for calculating predictive densities, instead using simulation-based bootstrap procedures to approximate them nonparametrically by sampling with replacement from regression residuals \citep{Efron79};
%


\item  broaden the information set from univariate to multivariate,  conditioning as well on natural covariates like sea ice thickness, surface air temperature, and radiative forcings, as for example in \cite{VARCTIC}.

\end{enumerate}

Alternatively, and of great interest, competitors may include large-scale structural dynamical climate models. That is, given a particular dynamical climate model, one could compare its ``model-based theoretical Figure  \ref{GlidePoint_SEP_2020}" to the ``data-based BPM Figure  \ref{GlidePoint_SEP_2020}" via skill score glide charts.  

In any event, comparison to the BPM  may prove useful for evaluating and selecting among various more sophisticated  sea ice models -- whether reduced-form or structural -- which are widely used to quantify the likely future evolution of Arctic conditions and their two-way interaction with economic activity.


\bibliographystyle{Diebold}
\addcontentsline{toc}{section}{References}
\bibliography{Bibliography}

@article{DGGC2022,
  title={Assessing and Comparing Fixed-Target Forecasts  of   Arctic  Sea Ice:  
	RMSE and Skill Score Glide Charts},
  author={Diebold, F.X. and G{\"o}bel, M. and Goulet Coulombe, P.},
  year={2022},
note={Manuscript in progress}

@article{StroeveEtAl2014,
author = {Stroeve, J. and Hamilton, L.C. and Bitz, C.M. and Blanchard-Wrigglesworth, E.},
title = {Predicting September Sea Ice: Ensemble Skill of the SEARCH Sea Ice Outlook 2008–2013},
journal = {Geophysical Research Letters},
volume = {41},
number = {7},
pages = {2411-2418},
doi = {https://doi.org/10.1002/2014GL059388},
year = {2014}
}

@article{Efron79,
author = "Efron, B.",
doi = "10.1214/aos/1176344552",
journal = "Annals of Statistics",
month = "01",
number = "1",
pages = "1-26",
publisher = "The Institute of Mathematical Statistics",
title = "Bootstrap Methods: Another Look at the Jackknife",
url = "https://doi.org/10.1214/aos/1176344552",
volume = "7",
year = "1979"
}

@article{DRice,
  title={Probability Assessments of an Ice-Free Arctic: Comparing Statistical and Climate Model Projections},
  author={Diebold, F.X. and Rudebusch, G.D.},
  year={2022},
journal={Journal of Econometrics},
note={in press.}
}

@article{Ing2003, title={Multistep Prediction in Autoregressive Processes}, volume={19}, DOI={10.1017/S0266466603192031}, number={2}, journal={Econometric Theory}, publisher={Cambridge University Press}, author={Ing, C.-K.}, year={2003}, pages={254–279}}

@article{VARCTIC,
  title={Arctic Amplification of Anthropogenic Forcing: A Vector Autoregressive Analysis},
  author={Goulet Coulombe, P. and G{\"o}bel, M.},
  journal={Journal of Climate},
note={\url{https://doi.org/10.1175/JCLI-D-20-0324.1}},
  year={2021},
pages={5523-5541}
}

@article{SIOinterim2020,
        author = {Meier, W. and U.S. Bhatt and J. Walsh and  R. Thoman and  P. Bieniek and  C.M. Bitz and  E. Blanchard-Wrigglesworth and  H. Eicken and  L.C. Hamilton and  M. Hardman and  E. Hunke and  T. Jung and  J. Kurths and  J. Little and  F. Massonnet and  J.E. Overland and  M. Serreze and  M. Steele and  J. Stroeve and  M. Wang and H.V. Wiggins},
        title = {},
        year = {2021},
        note = {\emph{2020 Sea Ice Outlook Post-Season Report} (Edited by  B. Turner-Bogren, L. Sheffield Guy, and S. Stoudt), \url{https://www.arcus.org/sipn/sea-ice-outlook/2020/post-season}}
}

@article{HamiltonStroeve2016,
author = {Hamilton, L.C. and Stroeve, J.},
title = {400 Predictions: the SEARCH Sea Ice Outlook 2008-2015},
journal = {Polar Geography},
volume = {39},
number = {4},
pages = {274-287},
year = {2016},
}

@Article{Hamilton2020,
	author = {Hamilton, L.},
	title = {1000 Predictions: What's New and What's Old in a Retrospective Analysis of the Sea Ice Outlook, 2008-2020},
	journal = {},
	year = 2020,
	volume = {},
pages={},
note={Presentation at American Geophysical Union Annual Meeting}
}

@ARTICLE{Fettereretal2017,
  author = {Fetterer, F. and Knowles, K. and Meier, W. and Savoie, M. and Windnagel, A.K.},
  title = {Sea Ice Index, Version 3, Dataset ID G02135},
note={Boulder, Colorado, USA. NSIDC: National Snow and Ice Data Center. \url{https://doi.org/10.7265/N5K072F8}, updated daily},
  year = {2017}
}

@article{hastie2009,
  note={\emph{The Elements of Statistical Learning}, Springer},
	year = 2009,
	author = {Hastie, T. and Tibshirani, R. and Friedman, J.}
}
%
%
%
%
%
%

\end{document}